\documentclass[aps,prb,twocolumn,groupedaddress,amsmath,amssymb]{revtex4}

\usepackage{longtable}
\usepackage{graphicx}
\usepackage{dcolumn}
\usepackage{bm}

\begin{document}


\title{Effect of biaxial strain and composition on vacancy mediated
  diffusion in random binary alloys:\\A first principles study of the
  Si$_{1-x}$Ge$_x$ system}


\author{Panchapakesan Ramanarayanan}
\email[]{panchram@stanford.edu}
\author{Kyeongjae Cho}
\affiliation{Department of Mechanical Engineering, Stanford University,
  Stanford, California 94305}
\author{Bruce M. Clemens}
\affiliation{Department of Materials Science and Engineering, Stanford
  University, Stanford, California 94305}


\nopagebreak

\begin{abstract}
We present the results of a systematic study using the density functional
theory (within the local density approximation) of the effects of biaxial
strain and composition on the self-diffusion of Si and Ge in Si$_{1-x}$Ge$_x$
alloys diffusing by a vacancy mechanism. The biaxial strain dependence of
the vacancy formation energy was reconfirmed with previous
calculations. The effect of biaxial strain on the interaction potential
energy between a substitutional Ge atom and a vacancy was calculated. The
interaction potential energy included not only the ground state energies
of the vacancy at different coordination sites from the Ge atom
but also the migration energy barriers to jump from one coordination site
to the adjacent.  These calculations were used to estimate the change in
the activation energy (due to biaxial strain) for the self-diffusion of Si
and Ge in Si by a vacancy mechanism.  The composition dependence of the
vacancy formation energy was calculated.  A database of {\em ab initio}
migration energy 
barriers for vacancy migration in different local environments was
systematically developed by considering the effect of the first nearest 
neighbor sites explicitly and the effect of the other sites by a mean field
approximation.  A kinetic Monte Carlo simulation based on the migration energy
barrier database was performed to determine the dependence (on the
composition) of the activation energy for the diffusion of Si and Ge in
Si$_{1-x}$Ge$_x$.  A detailed study of the variation of the correlation
factor with composition and temperature in Si$_{1-x}$Ge$_x$ was performed
using the results of the KMC simulation.  These analyses constitute
essential building blocks to understand the mechanism of vacancy mediated
diffusion processes at the microscopic level.
\end{abstract}


\maketitle


\section{\label{sec:intro}Introduction}
Silicon germanium technology is becoming increasingly popular in high
frequency, low power applications,\cite{ibmWWW} the principal reasons
being the advancement in
precision growth technologies\cite{meye92a} and the compatibility of
Si$_{1-x}$Ge$_x$ with the Si manufacturing processes along with such
properties of Si$_{1-x}$Ge$_x$ as the composition dependence of the band
gap, the strain dependence of the carrier mobility, and the increased
dopant solubility in Si$_{1-x}$Ge$_x$ compared to Si. The abrupt change in
the Ge concentration between the Si and the Si$_{1-x}$Ge$_x$ layers being a
functional necessity in these devices, a key materials issue is that of
interdiffusion in these layers.  There have been extensive studies of this
stress-coupled interdiffusion. 
\cite{aube02a,bari93a,iyer89a,prok95a,chan89a,scho91a,limY00a,limY02a}
Yet, a lot about the actual microscopic mechanisms behind these phenomena
remains to be understood.  Similarly the microscopic mechanisms
responsible for the growth, composition, and the shape of Si$_{1-x}$Ge$_x$
islands on Si, which find applications in areas like Si-based quantum
dots, are also not well understood.  From a technological standpoint,
understanding diffusion in Si$_{1-x}$Ge$_x$ is therefore important. 
From a scientific standpoint, the silicon germanium system presents
an ideal and clean system (without the complications introduced by
charged defect states) to further the theoretical understanding of
interdiffusion in random alloys in general.  Quoting from a recent paper,
\cite{zang01a} ``Theoretical treatments of self-diffusion in SiGe are
uncharted areas - and the effect of strain even more so.'' These reasons
have motivated us to perform a systematic and detailed first principles
study of the Si$_{1-x}$Ge$_x$ system. 

In their recent paper, Zangenberg {\em et al.} \cite{zang01a} have presented
their results of a systematic experimental study of the variation of Ge
self-diffusion in mono crystalline Si$_{1-x}$Ge$_x$ epi-layers as a
function of composition ($x$) and biaxial strain.  (A similar study has also
been reported by Strohm {\em et al.}.\cite{stro01a})  These works represent an
advancement over that presented by McVay and DuCharme \cite{mcva74a} in 1974,
which studied the composition dependent Ge diffusion in polycrystalline
Si$_{1-x}$Ge$_x$.  These results have been used in the past as an input to 
empirically
explain other experimentally observed phenomena.  For example, Baribeau
\cite{bari93a} used the Ge dependent diffusivity from Ref.~\onlinecite{mcva74a}
 to
numerically solve the one dimensional Fick's diffusion equation and
compared the results with those of the experimentally determined ones. 
Similarly, Aubertine {\em et al.} \cite{aube02a} have used the Ge dependent
diffusivity from Ref.~\onlinecite{zang01a} in a commercial numerical solver to
perform a similar comparison with their experiments to provide an
empirical explanation to the experimentally observed time dependent
interdiffusivity in Si/Si$_{1-x}$Ge$_x$ multilayers.  Thus, although these
results (of Refs.~\onlinecite{zang01a,mcva74a,stro01a}) have been valuable in
providing empirical insights into other phenomena, the reasons for these
behaviors themselves have not been queried into from a fundamental level,
to the best of our knowledge.  We are aware, however, of a recent paper by
Venezuela {\em et al.}\cite{vene02a} that has made an attempt in this 
direction. 

Our present work is a part of a project intended to develop a fundamental
understanding at the microscopic level ultimately, of the coupled strain
relaxation and interdiffusion phenomenon in Si/Si$_{1-x}$Ge$_x$
multilayers.  Towards this end, a fundamental understanding of the strain
and composition dependent diffusivity in Si$_{1-x}$Ge$_x$ as observed by
the experiments mentioned previously \cite{zang01a,stro01a,mcva74a} is an
essential prerequisite.  
Diffusion in SiGe has been postulated \cite{fahe89a} to be mediated 
by point defects:
vacancies, interstitials and by a point defect free mechanism - the
concerted exchange mechanism.  Very recently, another point
defect, which the authors \cite{goed02a} have termed the fourfold 
coordinated defect
(FFCD), has been suggested which could also be responsible for diffusion
in SiGe.  From their experimental observations, Fahey {\em et al.}
\cite{fahe89b} suggest that at 1050 $^\circ$C, the vacancy mechanism probably
contributes to 60\% - 70\% of the Ge diffusion in Si, the rest being due to
the interstitial mechanism.  We note that the results presented in
Ref.~\onlinecite{fahe89b} are for the diffusion of Ge in pure Si and so the
relative 
contributions of the different mechanisms could be different for systems
with different Ge concentrations.  However, because vacancies are among
the important contributors, they are the focus of this article. 

Density functional theory (DFT) calculations have played a significant
role in computational physics during the past few decades since the
theory's formal 
inception in the mid 1960s.  The unknown nature of the exchange
correlation functional and the inability to make progressively more
accurate approximations to the same (as would be possible, for example, in
a wave function based method), however, has been one of the main issues
concerning the practical application of the DFT.  The search for better
exchange correlation functionals is an active area of research in the
theoretical physics community.  The popular local density approximation
(LDA) and the more computationally expensive (but not necessarily more
accurate) 
generalized gradient approximation (GGA), unfortunately, have been unable
to reproduce experimentally observed values of activation energy of
diffusion in Si, the discrepancy \cite{souz98a} being as high as 1eV.  Quantum
Monte Carlo techniques, \cite{leun99a,foul01a} which circumvent the 
problem due to the exchange
correlation functional, are gaining popularity . 
However, because the LDA based DFT is definitely one of the most advanced
computational tools available for the systems of the size that we would
like to study, we have used it in this present study.  Because we are
aware of this discrepancy between the theoretical prediction and the
experimental values, and because we have restricted this present article
to only the vacancy mechanism, we prefer to refrain from making very
strong comparisons of our results with those of experiments, leaving such
comparisons to the future until we have resolved these outstanding issues. 
In spite of these limitations, we believe that our contribution is
significant for the following two reasons: (i) Our results can be viewed
as that of the behavior of a hypothetical random binary alloy system (with
the energetics provided by the LDA) diffusing by a vacancy mechanism. 
This is of basic scientific interest. (ii) The infrastructure that we
have developed in this present work can be reused with little effort once
accurate energetics becomes available.  This, along with similar analyses
for other diffusion mechanisms can then be used to directly compare/predict
experimental observations. 

\begin{figure}
\includegraphics[width=8.5cm]{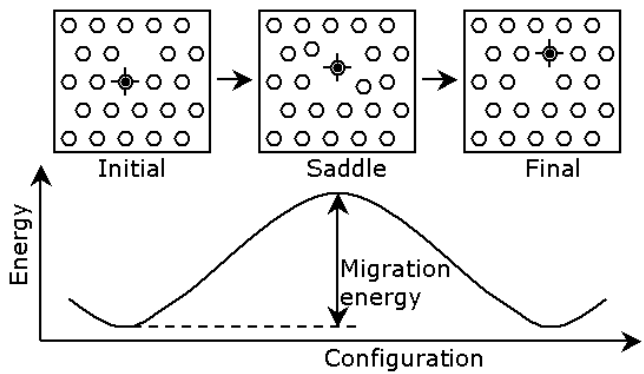}
\caption{\label{fig:unary_saddle} Migration energy for the motion of a
tracer (shown as the target symbol) in a unary system has a direct
  physical correspondence:
  difference between the saddle point and the ground state energies.}
\end{figure}

\begin{figure}
\includegraphics[width=5.5cm]{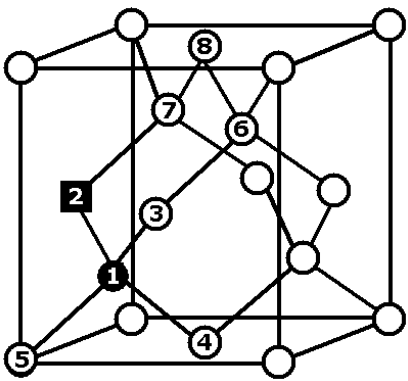}
\caption{\label{fig:diamond} Schematic of the Si structure. Open circles -
  Si; filled circle - Ge; filled square - vacancy.}
\end{figure}

\begin{figure}
\includegraphics[width=8.5cm]{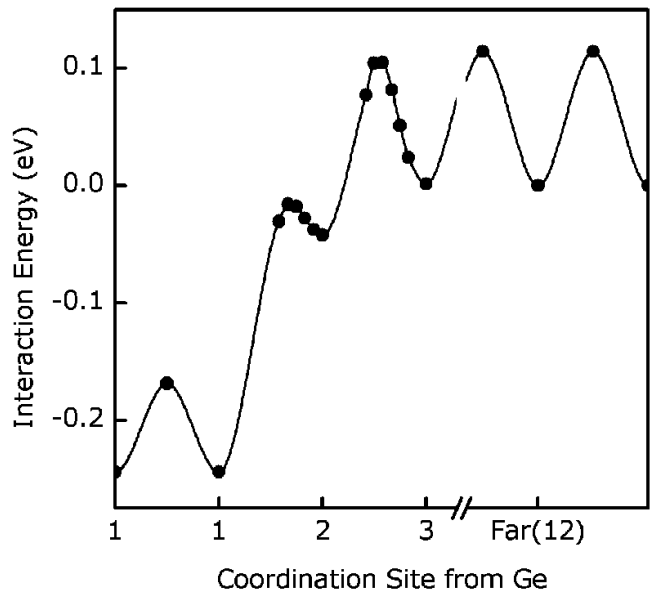}
\caption{\label{fig:GeV} 
Interaction potential energy (in eV) between a substitutional Ge
  atom and a vacancy as a function of vacancy position in relaxed Si from
  LDA calculations. (Lines are drawn as a guide to the eye.)}
\end{figure}

Because diffusion is a thermally activated process, diffusivity can be
characterized by a temperature independent term (the pre-exponential
factor, $D_0$) and a temperature dependent term (the exponential of the
activation energy, $E_a$, i.e., $\exp[-E_a/k_BT]$ where $k_B$ and $T$ 
are respectively
the Boltzmann's constant and the temperature). For a defect mediated
diffusion mechanism, the activation energy ($E_a$) is composed of the defect
formation energy and the rest, which we have termed as the
activation-minus-formation (AMF) energy.  We note that this has
traditionally been called as the migration energy. The reason we have
not called 
it the migration energy needs some explanation. We first consider the
case of a tracer self diffusion. Shown in Fig.~\ref{fig:unary_saddle}, for the
sake of illustration, is the motion of a tracer in a two dimensional
hexagonal lattice from an initial state to the final through the saddle
point. The migration energy in this case has a
very direct physical significance, namely, the energy difference between the
saddle point and the initial (ground) state. But when one considers anything
other than a unary system, for example the diffusion of a tracer Ge in Si,
one is unable to make a physically appealing correspondence to a
migration process as in the case of a tracer self diffusion in a unary
system. Specifically, considering the illustration shown in
Fig.~\ref{fig:diamond}, for the Ge atom (filled circle) to be effectively
displaced from its current position (labelled 1), the vacancy (filled
square) has to move from its current position (labelled 2) to atleast
the third coordination site from Ge (labelled 8) and return by another
path (for example through sites labelled 6 and 3). As shown in
Fig.~\ref{fig:GeV}, there are different energy barriers
to get to different configurations: the barrier for the vacancy to get
from the first coordination site (from Ge) to the second is different from
that to get from the second to the third. The contribution to the
activation energy (other than the effective defect formation energy) is
not only from the complex collective 
action of all these different migration barriers but also, in this particular
pair diffusion model, due to the energy required for the vacancy to get to
the third coordination site, for example, so that it can return to the 
Ge atom from
a different direction thereby causing a net motion of the Ge atom. 
These complex collective actions manifest in different forms, for example,
as a temperature dependent correlation factor. Therefore, we felt the
need to make the distinction from the term: migration energy.  The AMF energy 
equals the migration energy in the microscopic sense only for the case of 
the tracer self-diffusion in a unary system. Although previous reports
\cite{dunh95a,pank97a,sugi92a} have suggested different measures of an
effective migration barrier for these cases, we remark that we do not find
any of them physically enlightening. 

In order to study the effect of biaxial strain and composition on the
diffusivity, one needs to study their effect on these three parameters,
viz., the pre-exponential factor ($D_0$), the vacancy formation energy, and the
AMF energy.  Several previous first principles calculations
\cite{anto90a,anto98a,wind99a,sugi92a} have reported on the strain
dependence of 
vacancy formation energy.  In this article, we present our results where
we have reconfirmed the biaxial-strain dependence of the vacancy formation
energy.  Though we have seen general theoretical treatments
\cite{dawM01a,dede78a} of the effect
of strain on defect diffusion, we have not come across
any reference reporting on the {\em ab initio} based calculation of the
variation of the AMF energy (or the effective migration energy as it is
generally known) as a function of strain.
In this article, we have 
calculated the biaxial strain dependence of the interaction potential energy
between a substitutional Ge atom and a vacancy.  The interaction potential
energy included not only the ground state energies of the vacancy at
different coordination sites from the Ge atom but also the migration
energy barriers for jumps from one coordination site to the adjacent.  We
have used these calculations to estimate the change in the activation
energy (due to biaxial strain) for the self-diffusion of Si and Ge in Si
by a vacancy mechanism.  We then present our calculations of the Ge
concentration dependence of the vacancy formation energy where we have
used the classic Boltzmann factor enhancement of the probabilities.  We
note that Venezuela {\em et al.}\cite{vene02a} have used an approach similar
to ours in their 
recent paper.  Earlier theoretical or numerical studies
\cite{dunh95a,math82a,yosh71a,bune00a} have 
only reported on the analyses of the concentration dependence of diffusivity
in the low concentration regime, typically those corresponding to dopant
concentrations.  We have not found
theoretical or numerical treatments of the concentration effects on
diffusivity for higher solute concentrations like those found in
Si$_{1-x}$Ge$_x$ alloy systems.  We present the variation of the
AMF energy as a function of Ge concentration,
which we have obtained from kinetic Monte Carlo (KMC) simulations
using a migration energy barrier database calculated from first
principles.  The KMC simulations also 
enabled us to study the variation of the correlation factor as a function
of Ge concentration and temperature.  Such a study provides useful insight
into the vacancy mediated diffusion mechanism in a random binary alloy
arranged in a tetrahedral geometry. 

This article is organized as follows:  In section~\ref{sec:theory}, we
present the 
theoretical details of our calculations and the computational details of
our simulations.  In section~\ref{sec:results}, we discuss our main
results.  In section~\ref{sec:summary}, we conclude the article with a
summary of the present work and also comment on the limitations of this work.

\section{\label{sec:theory}Theoretical and Computational Details}
In this section, we present the theory behind our calculations and the
main computational details of our simulations.
\subsection{\label{sec:DFTTheo}First principles calculations}
Our first principles calculations were performed using the plane-wave
ultrasoft pseudopotential code VASP \cite{VASP1,VASP2,VASP3,VASP4} within
the local density 
approximation (LDA).  A 64-atom supercell with a kinetic energy cut-off of
10 Hartree, and a $2^3$ Monkhorst-Pack \cite{monk76a} {\bf k}-point
sampling was used. 
Electronic energy convergence of up to $2.7 \times 10^{-5}$ eV was used and the
structures were relaxed until the maximum force on any atom was less than
0.015 eV/{\r{A}}.  Saddle point configurations were determined using the nudged
elastic band (NEB) method. \cite{NEB}  Optimized Si and Ge lattice constants
(computed by fitting the total energy vs. the supercell volume to
Murnaghan's \cite{murnFN,murn44a} equation of state) were found to be 5.39 {\r{A}} and
5.62 {\r{A}} respectively.  The vacancy formation energy and the vacancy
formation volume in Si (Ge) were found to be 3.31 (1.88) eV, and -0.059
(-0.195)$\Omega$  respectively where $\Omega$ represents the volume of a
Si (Ge) atom. 
(We recall that the vacancy formation volume is the sum of the relaxation
volume (Si: -20.73 {\r{A}}$^3$; Ge: -26.56 {\r{A}}$^3$) and the atomic
volume (Si: 19.57 {\r{A}}$^3$; Ge: 22.23 {\r{A}}$^3$).)  These values are
expectedly comparable to other recent first 
principle calculations. \cite{pusk98a,merc98a,dabr00a,fazz00a}  Because
of the low Si-vacancy formation volume and because of Si$_{1-x}$Ge$_x$ being a
model random alloy, \cite{giro91a,gegeFN}
the lattice parameters for
Si$_{1-x}$Ge$_x$ were chosen by a simple rule of mixtures.  Lattice
parameters so 
chosen were assumed to correspond to a strain relaxed state. Unless
otherwise mentioned, all our calculations were done in such a strain
relaxed state.
\subsection{\label{sec:KMCTheo}Kinetic Monte Carlo simulations}
The diamond lattice was generated in the computer memory by a
four-dimensional integer array.  The first three indices were used to
reference the location (i.e., X,Y,Z ``coordinates'') of the cubic unit cell.
The last index was used to reference the particular atom among
the eight in the unit cell referred to by the first three indices.
Because Si$_{1-x}$Ge$_x$ forms a random alloy,\cite{giro91a,gegeFN} each
member of this array was randomly designated as either a Si atom or a Ge
atom and one randomly chosen member was designated as the vacancy.  The
relative numbers of the Si and Ge atoms were so chosen that the required
composition was obtained.  The displacement and the number of jumps
performed by each of these atoms were recorded through out the
simulation.  Periodic boundary conditions were used so that an atom
hopping out of one side reenters the system from the opposite side,
essentially simulating an infinite system. 

Each KMC move consisted of the following five steps: (i) Obtaining the
rates $r_i$ for the possible final configurations starting from the current
configuration as the initial configuration. (ii) Generating a
pseudo-random number $\gamma \in (0,1]$. (iii) Advancing the clock
\cite{bort75a} by $-\ln(\gamma)/\sum_ir_i$.
(iv) Reconfiguring the system into one of the final configurations based
on the random number generated in step (ii). (v) Updating the displacement
and the number of jumps of the vacancy and the atoms that have moved.
We refer the reader to the original KMC paper by Voter \cite{vote86a} for
details on the theory of the kinetic Monte Carlo algorithm. 

The supercell comprised of $50 \times 50 \times 50$ cubic unit cells each
containing eight lattice sites making up one million lattice sites.
Random alloys of 
Si$_{1-x}$Ge$_x$ with the concentration of Ge ($x$) varying from 0 to 1
were used to study the effect of composition.  A single vacancy was used.
(We note that the presence of one vacancy in a $50 \times 50 \times 50$
super cell, 
which we are forced to use due to computational limitations, results in an
extremely high concentration of vacancy compared to the real
Si$_{1-x}$Ge$_x$ system.)  Three different random distributions of Ge
atoms were used for each composition and three different random number
sequences were used for each distribution, thus making up nine samples for
each composition.  The results were averaged over all the nine samples.
A billion vacancy hops were performed for each case. The scatter in
the results among these nine samples was found to be extremely low.
\subsection{\label{sec:dEfXGeTheo}Effective vacancy formation energy calculation}
In this subsection, we explain how we use the classic Boltzmann factor to
calculate the effective vacancy formation energy in the Si$_{1-x}$Ge$_x$
alloy (where $x$ denotes the atomic fraction of Ge).  The strength of the
interaction between a vacancy and a Ge atom is expectedly dependent on
their relative positions. The interaction of the vacancy with the Ge atoms
which are first nearest neighbors to the vacancy is stronger than with
those which are second nearest neighbors. This second nearest neighbor
interaction in turn is stronger than between those which are further
away. We have therefore 
chosen three different forms to represent these three different
interactions.  For the strongest interaction, we define a function $F$.  We
denote as $F(b)$ the drop in energy of the system when a vacancy is
surrounded by $b$ Ge atoms at the first nearest neighbor position to the
vacancy. (For the Si structure, $b$, of course, ranges from zero through
four.) For the interaction between the vacancy and Ge atoms that are at 
the second nearest neighbor positions to the vacancy, we use a linear
expression for the 
drop in the energy with the number of Ge atoms in the second nearest
neighbor position.  We denote the constant of proportionality i.e., the
drop in energy of the system for each Ge atom in the second nearest
neighbor position as $S$.  Because the interaction between the vacancy and
Ge atoms that are present beyond the second nearest neighbor positions is
comparatively weak, we consider their effect through a mean field
correction factor: $M$.  $M$ is the difference between the energy of a system
with a vacancy whose first and second nearest neighboring positions are
occupied by Ge atoms and all other positions are occupied by Si atoms and
that of a system with a vacancy with Ge atoms in all the positions. We
denote as $E(n,b,x)$ the vacancy formation energy in a Si$_{1-x}$Ge$_x$
system (with a Ge concentration of $x$). Here, $n$ denotes the total number of 
Ge atoms in the first and second nearest neighbor positions to the
vacancy and $b$ denotes the number of Ge atoms that are in the first
nearest neighbor positions to the vacancy. We then obtain the following
expression for $E(n,b,x)$ in terms of the 
vacancy formation energy in pure Si ($E^{Si}_{V_f}$) and the quantities
$F$, $S$, and $M$ defined above:
\begin{eqnarray}
E(n,b,x) = E^{Si}_{Vf} - F(b) - (n-b)S - Mx
\label{eq:Enfx}
\end{eqnarray}
If the distribution of Ge atoms is unaffected by vacancies, then, the
probability $\tilde{p}(n,b,x)$ of a vacancy being surrounded by $n$ Ge atoms, 
$b$ of which are
first nearest neighbors to the vacancy and the rest ($n-b$) are second
nearest neighbors is calculated in the following manner using the binomial
Bernoulli distribution from elementary probability theory:  (Note: There
are 4 first nearest neighbor sites and 12 second nearest neighbor
sites in the diamond lattice.)  The probability of a vacancy being
surrounded by $b$ Ge atoms in the first nearest neighbor position is
${4 \choose b} x^b (1-x)^{4-b}$
(where $x$ is the concentration of Ge).  The probability of a vacancy being
surrounded by $(n-b)$ Ge atoms in the second nearest neighbor position is
${12 \choose n-b} x^{n-b} (1-x)^{12-(n-b)}$.
The required probability $\tilde{p}(n,b,x)$ is therefore the product of the 
above two which simplifies to the following expression:
\begin{eqnarray}
\tilde{p}(n,b,x) = {4 \choose b} {12 \choose n-b} x^n (1-x)^{16-n}
\end{eqnarray}
The interaction between the Ge and the vacancies, however, affects their
distribution. The probability $p(n,b,x,T)$, which takes this interaction
into account, is obtained by multiplying $\tilde{p}(n,b,x)$ by the
Boltzmann factor corresponding to the energy drop because of this
interaction. We thus obtain the following expression for $p(n,b,x,T)$:
\begin{eqnarray}
p(n,b,x,T) = \tilde{p}(n,b,x) \exp[(F(b)+ (n-b)S)/k_BT]
\label{eq:pnbxT}
\end{eqnarray}
where $k_B$ is the Boltzmann's constant and $T$ is the temperature. We
then express the effective vacancy formation energy, $\langle E_f(x,T)
\rangle$, as an average of the vacancy formation energies in the different
environments weighted by their corresponding probabilities:
\begin{widetext}
\begin{eqnarray}
\left\langle E_f(x,T) \right \rangle &=&
\frac{1}{z}
\left[ \sum_{n=0}^4
\left( \sum_{b=0}^n
p(n,b,x,T) \times E(n,b,x)\right)
+ \sum_{n=5}^{16}
\left( \sum_{b=0}^4
p(n,b,x,T) \times E(n,b,x) \right)
\right]
\label{eq:EfxT}
\end{eqnarray}
\end{widetext}
where $z$ is like the partition function:
\begin{eqnarray}
z = \sum_{n=0}^4 \left( \sum_{b=0}^n p(n,b,x,T) \right) +
\sum_{n=5}^{16} \left( \sum_{b=0}^4 p(n,b,x,T) \right)
\label{eq:z}
\end{eqnarray}
We make the following two clarifications: (i)We have not included the 
mean field correction term $M$ in the expression for the Boltzmann factor 
in Eq.~(\ref{eq:pnbxT}) because, $M$ being independent of $n$ or $b$,
is factored out of the numerator and the denominator ($z$) in the 
expression for $\langle E_f(x,T) \rangle$ (Eq.(~\ref{eq:EfxT})) even if it is
included.
(ii) We have two terms in the RHS of Eqs.~(\ref{eq:EfxT}) and (\ref{eq:z})
for the following simple reason: The number of
first nearest neighbor Ge atoms ($b$) can be at most equal to the total
number of Ge atoms in the first and the second nearest neighbor positions
($n$) when $n$ is less than or equal to four.  This is the first term on the
RHS.  The variable $b$ can be at most equal to four when $n$ is greater
than four.  This is the second term.
\subsection{\label{sec:CorrTheo}Theoretical calculation of correlation
factor }
The correlation factor ($f$) is defined \cite{mann68a} as the ratio of the
actual diffusion 
coefficient to the uncorrelated diffusion coefficient under the assumption
that all the jumps are statistically independent of one another. The
correlation factor provides a lot of insight into the microscopic
mechanism of diffusion.
In this sub-section, we give a brief outline of how the correlation factor
is computed.  We refer the reader to Ref.~\onlinecite{mann68a} for further
details. The correlation factor for the diffusion of a single impurity
atom in a cubic crystal by the vacancy mechanism can be calculated using
the expression:  
\begin{eqnarray}
f=\frac{1+\langle\cos\theta\rangle}{1-\langle\cos\theta\rangle}
\label{eq:fcorrdefn}
\end{eqnarray}
Here, $\langle\cos\theta\rangle$, which denotes the average of the cosine
of the angle between 
successive impurity jumps, can be evaluated as $T_j\cos\theta_j$. $T_j$ is
the probability for the impurity to jump to the $j^{\text{th}}$
configuration. $\theta_j$ is 
the angle formed between the current impurity jump direction and the
impurity jump direction leading
to the $j^{\text{th}}$ configuration. There is the implicit sum over the
repeated 
index $j$.  In the case of the diamond structure, $j$ ranges over the four
first nearest neighbors; i.e., the $j^{\text{th}}$ configuration results
when the 
impurity jumps to the $j^{\text{th}}$ first nearest neighbor.  Referring to
Fig.~\ref{fig:diamond}, 
where the destination configurations (i.e., first nearest neighbors to
impurity) are denoted by the numbers 2 through 5, the probabilities $T_2$
through $T_5$ have been calculated by including jump sequences up to five
vacancy hops \cite{correlFN}
in the following manner:  We denote 
respectively by $\nu_I$, $\nu_H$, $\nu_F$, and $\nu_B$, the jump rates for
the following 
vacancy jump processes: (i) Vacancy exchanges positions with the
impurity atom. (ii) Vacancy exchanges positions with the host atom without
either breaking or forming a bond with the impurity atom. (iii) Vacancy
exchanges positions with the host atom and in the process {\em forms} a bond
with the impurity atom. (iv) Vacancy exchanges positions with the host atom
and in the process {\em breaks} a bond with the impurity atom.  (We explain the
procedure for obtaining the values of the various jump rates ($\nu_I$,
$\nu_H$, $\nu_F$, and $\nu_B$) from our LDA calculations in
Sec.~\ref{sec:StrnRslt}.)  We denote as $R(j,k)$ the 
probability for the impurity to jump to the $j^{\text{th}}$ first nearest
neighbor 
position as a result of the vacancy performing $k$ hops. (For example
(referring to Fig.~\ref{fig:diamond}), one of the ways in which the
impurity atom can 
jump to the first nearest neighbor position denoted as 2 as a result of
the vacancy performing three hops would be the following jump sequence of
the vacancy: 2 to 7 followed by 7 to 2 followed by 2
to 1.) 
Using elementary probability theory, the various $R(j,k)$'s can be computed to
obtain:
\begin{eqnarray}
R(2,1) &=& \frac{\nu_I}{\nu_I + 3\nu_B} \\
R(2,3) &=& 3 \times \frac{\nu_B}{\nu_I + 3\nu_B} 
\times \frac{\nu_F}{\nu_F +3\nu_H} 
\times \frac{\nu_I}{\nu_I + 3\nu_B} \\
R(2,5) &=& 9 \times \frac{\nu_B}{\nu_I + 3\nu_B} 
\times \frac{\nu_H}{\nu_F +3\nu_H} 
\times \frac{1}{4}\nonumber\\
&&\times \frac{\nu_F}{\nu_F + 3\nu_H}
\times \frac{\nu_I}{\nu_I + 3\nu_B}\\
R(3,5) &=& R(4,5) = R(5,5) =\nonumber\\
&&2 \times \frac{\nu_B}{\nu_I + 3\nu_B}
\times \frac{\nu_H}{\nu_F +3\nu_H} 
\times \frac{1}{4}\nonumber\\
&&\times \frac{\nu_F}{\nu_F + 3\nu_H}
\times \frac{\nu_I}{\nu_I + 3\nu_B}
\end{eqnarray}
The $R(j,k)$'s not listed above are all zero. $T_j$ is then obtained by simply summing
$R(j,k)$ over $k$ from one through five.\cite{correlFN} We obtain the following expressions,
in terms of the various $R(j,k)$'s, for the probabilities $T_2$ through $T_5$: 
\begin{eqnarray}
T_2 &=& R(2,1) + R(2,3) + R(2,5) \\
T_3 &=& T_4 = T_5 = R(3,5)
\end{eqnarray}
Also, from Fig.~\ref{fig:diamond}, $\cos\theta_2 = -1$ and 
$\cos\theta_3 = \cos\theta_4 = \cos\theta_5 = 1/3$. 

\subsection{\label{sec:CorrKMCTheo}Calculation of the correlation factor from
the KMC simulation results} 
The procedure for calculating the correlation factor outlined in
Sec.~\ref{sec:CorrTheo} is valid only for a single impurity atom migrating by
a vacancy mechanism.  Certain symmetry requirements, which were implicitly
used in the formulae presented there, are violated
at higher impurity concentrations.  In this subsection, we
explain the procedure for calculating the correlation factor that is valid
for any impurity composition, as long as there are a sufficient number of
atoms to obtain a good statistical average.  This procedure is a
straightforward interpretation of the definition of the correlation factor
as applied to the results from the KMC simulation. 

From the definition of the correlation factor as the ratio of the actual
diffusivity to the uncorrelated diffusivity and from the definition of the
diffusivity as the ratio of the mean squared displacement 
$\langle X^2 \rangle$ to $6\tau$, where $\tau$ 
is the time taken for the motion in the limit as $\tau$ tends to zero, we
obtain the correlation factor to be the ratio of the actual mean squared
displacement to the mean squared displacement when the motion is
uncorrelated.  Symbolically,
\begin{eqnarray}
f = 
\frac{\langle X^2 \rangle_{\text{actual}}}{\langle X^2 \rangle_{\text{random}}}
\end{eqnarray}
From the random walk analysis, 
$\langle X^2 \rangle_{\text{random}} = \langle N \rangle \lambda^2$, 
where $\langle N \rangle$ is the mean
number of jumps and $\lambda$ is the single jump distance.  For the
diamond structure, $\lambda = 0.25\sqrt{3}$ (in units of the unit cell
dimension) and so we obtain the correlation factor to be
\begin{eqnarray}
f = 
\frac{\langle X^2_x \rangle + \langle X^2_y \rangle 
+ \langle X^2_z \rangle}{3 \times (0.25)^2 \times \langle N \rangle}
\end{eqnarray}
From the output of the KMC simulation,
which has the displacements and the number of jumps of each atom, the mean
squared displacements ($\langle X^2_x \rangle$, $\langle X^2_x \rangle$, 
$\langle X^2_x \rangle$) and the mean number of jumps ($\langle N
\rangle$)
can be calculated by averaging the quantities over all atoms of the same
type (Si or Ge).  The 
correlation factor can thus be calculated in a straightforward manner from
the KMC simulation results.

\section{\label{sec:results}Results and Discussion}
\subsection{\label{sec:StrnRslt}Biaxial strain dependence of the
activation energy} 

The change in vacancy formation energy due to biaxial strain is computed
as \cite{aziz97a} $(-2/3)V_r\mu$ where $V_r$ is the relaxation volume and
$\mu$ is the 
biaxial modulus of Si. From our LDA calculations, the relaxation volume
accompanying the formation of a vacancy in Si is -20.73 {\r{A}}$^3$.
(The relaxation volume in this case is approximately -1.06 times the
atomic volume of Si.)
The biaxial modulus\cite{kuzn99a} of Si is 190.48 GPa.  We
therefore find the change in vacancy formation energy due to equi-biaxial
strain to be 16 eV/unit strain. 

\begin{figure}
\includegraphics[width=8.5cm]{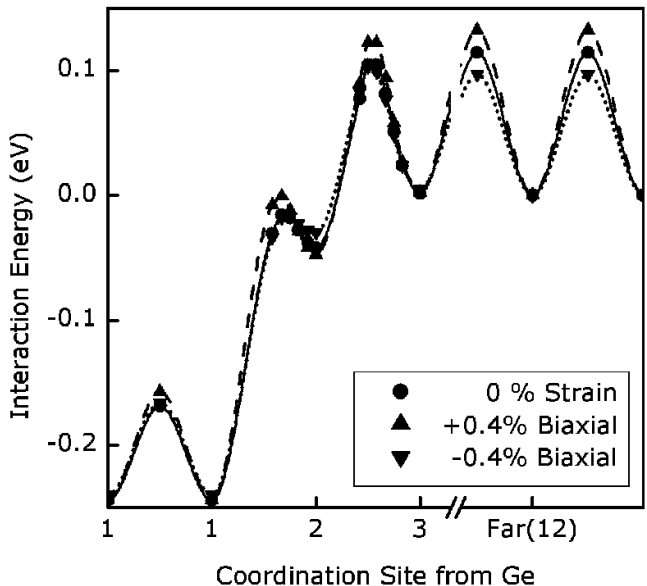}
\caption{\label{fig:GeVStrn} 
  Interaction potential energy (in eV) between a substitutional Ge
  atom and a vacancy as a function of vacancy position in (a) relaxed Si
  (circles connected with solid 
  line) (b) 0.4\% tensile biaxially strained Si (upward triangles connected
  with dashed line) (c) 0.4\% compressively biaxially strained Si (downward
  triangles connected with dotted line) from LDA calculations. The
  energies of all the systems when the vacancy and the Ge are far apart
  have been normalized to zero. (Lines are drawn as a guide to the eye.)}
\end{figure}

The interaction potential energies of a vacancy with a substitutional Ge
atom for (a) relaxed, (b) 0.4\% tensile equi-biaxially strained, and (c)
0.4\% compressively equi-biaxially strained systems are as shown in
Fig.~\ref{fig:GeVStrn}. (The interaction potential energy for the relaxed
system shown in Fig.~\ref{fig:GeVStrn} is exactly same as that shown in 
Fig.~\ref{fig:GeV}. It has been shown separately in Fig.~\ref{fig:GeV} for
the sake of clarity and has been shown in Fig.~\ref{fig:GeVStrn} for the
sake of making comparisons with the strained systems.)
We make the following comments and observations with
reference to Figs.~\ref{fig:GeV} and \ref{fig:GeVStrn}: 
(i) Our first principles calculations indicate that a biaxial tension 
(compression) of 0.4\% causes the third dimension to contract (expand) by
0.39\% (0.46\%) in reasonable agreement with that predicted by linear
elasticity theory \cite{wall86a} which gives a contraction (expansion) of
0.31\% (0.31\%).  To maintain consistency, the interaction potential energy
calculations were made using the dimensions obtained from our LDA
computations. (ii) From Fig.~\ref{fig:GeV} we see that the migration
barrier for the vacancy to exchange positions with a Si atom far away from
a Ge atom is 0.11 eV. This is as expected, owing to the similarity of the
calculation technique, in reasonable agreement with 
Nelson {\em et al.} \cite{nels98a} who report a value of 0.18 eV from
their LDA calculations. (iii) The asymmetric location of the saddle points
between the 1st and the 2nd coordination sites, and the 2nd and the 3rd
coordination sites is due to the weaker nature of the Si-Ge bond compared
to that of the Si-Si bond. (iv) From Fig.~\ref{fig:GeV} we see that the
binding energy of the vacancy to the Ge atom with the vacancy being at the 
$n^{\text{th}}$ coordination site from Ge is 0.24 eV, 0.04 eV, and less
than 0.002 eV 
respectively for $n = 1$, $2$ and $3$.  Therefore, the vacancy is practically
bound to Ge only if it is at a nearest neighbor site to Ge.  The strength
and the range of interaction between a vacancy and Ge is quite weak unlike
those between a vacancy and a dopant atom such as arsenic \cite{pank97a} or
phosphorous \cite{nels98a}.  This difference in the intensity and the extent of
the interaction and the difference in the typical concentration of Ge in
Si$_{1-x}$Ge$_x$ alloys compared to dopant concentrations suggests that the
diffusion of Ge will not be dominated by the pair diffusion mechanism,
which is the accepted \cite{yosh71a,pank97a} dominant mechanism of dopant
diffusion diffusing by the vacancy mechanism.  Rather, the vacancy by
randomly moving through the crystal randomly displaces Ge atoms whenever
it meets one, thereby causing diffusion.  It does not form as strong a
pair with the Ge atom as, for example, it does with a phosphorous or an
arsenic atom.
(v) From the interaction potentials (Fig.~\ref{fig:GeVStrn}), we find that
the barrier for the Si-V 
jump (far from a Ge atom) changes by 4eV/unit equi-biaxial strain and the
barrier for the Ge-V jump (at very low Ge concentration) changes by
2eV/unit equi-biaxial strain.  

\begin{figure}
\includegraphics[width=8.5cm]{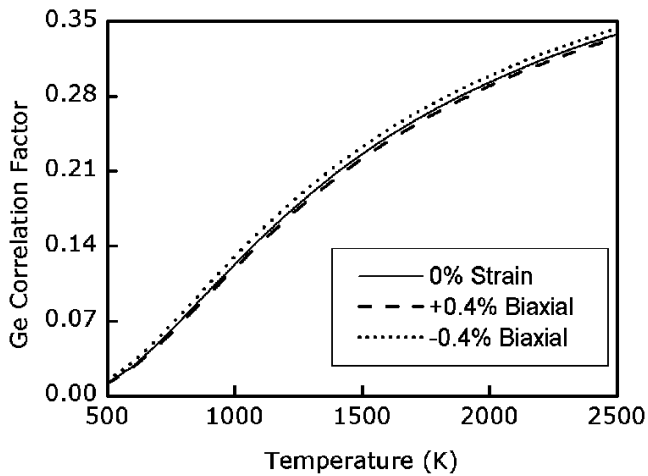}
\caption{\label{fig:GeCorrTheo} 
Theoretical calculation of the correlation factor for the
  diffusion of Ge in (a) relaxed Si (solid line) (b) 0.4\% tensile
  biaxially strained Si (dashed line) (c) 0.4\% compressively biaxially
  strained Si (dotted line) as a function of temperature. The correlation
  factors are seen to approach a high temperature limit of 0.5, the
  theoretical value for a tracer diffusion in a diamond structure.}
\end{figure}

\begin{figure}
\includegraphics[width=8.5cm]{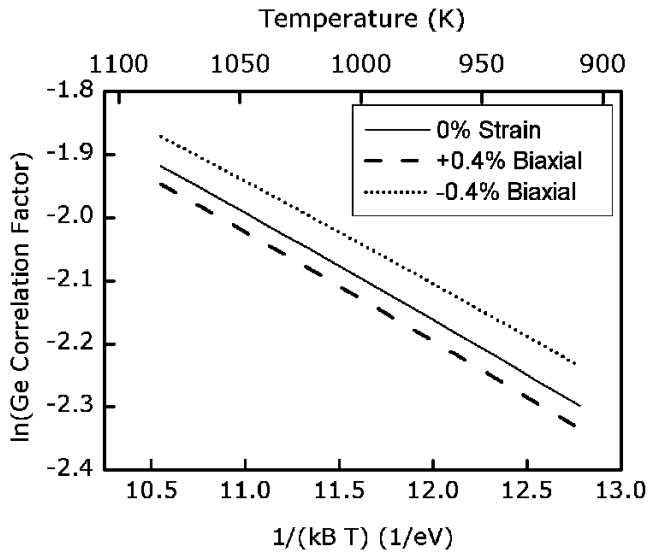}
\caption{\label{fig:GeCorrTheoArr} 
An Arrhenius type plot of the correlation factor for the 
  diffusion of Ge in (a) relaxed Si (solid line, $E_a^{corr}$ = 0.168 eV)
  (b) 0.4\% tensile biaxially strained Si (dashed line, $E_a^{corr}$ = 0.171
  eV) (c) 0.4\% compressively biaxially strained Si (dotted line,$E_a^{corr}$
  = 0.161 eV) to extract the activation energy corresponding to the
  correlation factor ($E_a^{corr}$).}
\end{figure}

The Ge-V interaction potential from Fig.~\ref{fig:GeVStrn} can be used to
calculate the correlation factor for Ge diffusion as outlined in 
Sec.~\ref{sec:CorrTheo}. From the transition state theory (TST), the
jump rates $\nu_I$, $\nu_H$, $\nu_F$ and $\nu_B$ mentioned in 
Sec.~\ref{sec:CorrTheo} can be calculated as:
\begin{eqnarray}
\nu_I &=& \nu_0 \exp[{-(E_{xs}-E_1)/k_BT}] \\
\nu_H &=& \nu_0 \exp[{-(E_{fs}-E_f)/k_BT}] \\
\nu_F &=& \nu_0 \exp[{-(E_{12}-E_2)/k_BT}] \\
\nu_B &=& \nu_0 \exp[{-(E_{12}-E_1)/k_BT}]
\end{eqnarray}
where $\nu_0$ denotes the lattice vibrational frequency
(which, to a first order approximation we have assumed to be a constant);
$E_1$, $E_2$ and $E_f$ respectively denote the energy of the system when
the vacancy is at 
the first nearest neighbor site to the Ge atom, second nearest neighbor
site to the Ge atom and far away from the Ge atom; $E_{xs}$, $E_{fs}$ and 
$E_{12}$ respectively denote saddle point energies for the vacancy and Ge to
exchange positions, for the vacancy and Si to exchange positions far away
from a Ge atom, and for the vacancy to move between the first and the
second nearest neighboring positions of the Ge atom.
Fig.~\ref{fig:GeCorrTheo} plots the
variation of the correlation factor for Ge diffusion 
as a function of temperature. (As we have noted previously,\cite{correlFN}
the correlation factor 
approaches the theoretical limit of 0.5 at high
temperatures.) In Fig.~\ref{fig:GeCorrTheoArr} we show an Arrhenius plot
of the same and 
we extract the activation energy for the strain free, 0.4\% biaxial
tensile and 0.4\% biaxial compressive cases to be 0.168 eV, 0.171 eV, and
0.161 eV respectively.  The activation energy associated with the
correlation factor for Ge diffusion in Si at very low Ge concentrations
therefore changes by approximately 1 eV/unit strain.  

Combining the results of the vacancy formation energy change due to
biaxial strain with 
the migration barrier energy change and the correlation factor activation
energy change, we estimate the following values for the effect of
equi-biaxial strain on the diffusion-activation energy: 20 eV/unit strain
for Si self diffusion in Si, 17 - 20 eV/unit strain for Ge self diffusion
in Si.  
At this point, we would like to quote two experimentally
determined values and one empirically fitted value for the change in
activation energy for Ge diffusion due to biaxial strain.  The
experimentally determined values are due to 
Cowern {\em et al.}\cite{cowe96a} and 
Zangenberg {\em et al.}\cite{zang01a} who report a value of 18 eV/unit
strain and 
160 $\pm$ 40 eV/unit strain respectively.  
Aubertine {\em et al.}\cite{aube02a}, who 
use the strain dependence of the activation energy as a tunable parameter
in their empirical model, report that they are best able to reproduce
their experimental data if they set this parameter to be 19eV/unit
strain. 

\subsection{\label{sec:dEfXGeRslt}Ge concentration dependence of the vacancy
formation energy} 

The typical concentration of Ge in SiGe films in device structures is 10\%
- 30\%, which is several orders of magnitude larger than the typical dopant
concentration ($10^{16}$ - $10^{18}$ per cm$^3$).  One therefore needs to
consider the 
effect of Ge concentration on the vacancy formation energy (and hence the
vacancy concentration) in the system in 
the manner explained in Sec.~\ref{sec:dEfXGeTheo}.  From straightforward
LDA based 
calculations, we obtain the following energetics of the SiGe-vacancy
system:  The energy of the system drops by 0.24, 0.45, 0.6, and 0.83 eV
when the vacancy is surrounded respectively by 1, 2, 3 and 4 Ge atoms.
Similar figures have been previously reported.\cite{bogu99a,vene02a}
The energy of the system drops by 0.04 eV for every second
nearest neighbor Ge atom to the vacancy.  The energy of the system with Ge
in all the second nearest neighbor positions of the vacancy and Si
everywhere else is higher than the energy of a system with a vacancy in
unary Ge by 0.12 eV.  In terms of the notations used in
Sec.~\ref{sec:dEfXGeTheo}, $F(0)=0$ eV, $F(1) = 0.24$ eV, $F(2)=0.45$ eV,
$F(3)=0.6$ eV, $F(4) = 0.83$ eV, $S = 0.04$ eV, and $M = 0.12$ eV.
The attractive interaction between the Ge atoms and the vacancy
causes the equilibrium vacancy concentration to be larger in regions of
high Ge concentration.  This lowers the effective vacancy formation energy
in Si$_{1-x}$Ge$_x$ compared to a uniform (regionally unbiased) random
distribution of vacancies. 

\begin{figure}
\includegraphics[width=8.5cm]{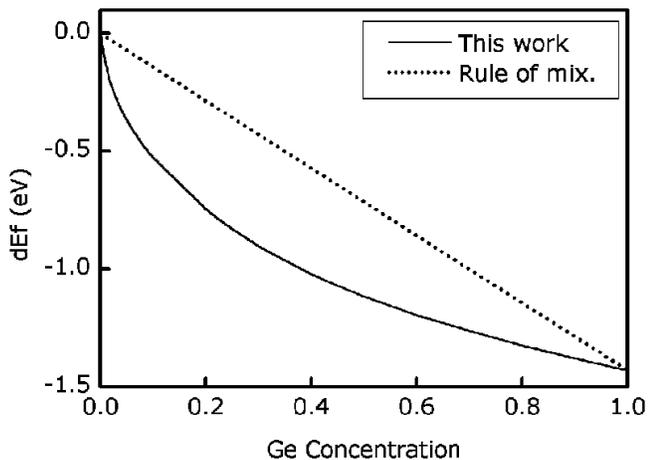}
\caption{\label{fig:dEfTheo} 
Solid line shows the change in the vacancy formation energy in
  Si$_{1-x}$Ge$_x$ from that in pure Si 
($dEf = \left\langle E_f(x,T) \right\rangle - E^{Si}_{V_f}$) as a function
  of Ge concentration 
  ($x$) calculated at 1000K by taking into account the attractive
  interaction of the vacancy with the Ge atoms. Dotted line shows the same
  quantity obtianed by a simple rule of mixtures.}
\end{figure}

Using the theory explained in Sec.~\ref{sec:dEfXGeTheo}, we have plotted in
Fig.~\ref{fig:dEfTheo} the change in the vacancy formation energy from the
vacancy formation energy in Si 
($\left\langle E_f(x,T) \right\rangle - E^{Si}_{V_f}$) as a function of Ge 
concentration calculated at 1000K (solid line).  Also plotted in
Fig.~\ref{fig:dEfTheo} is 
the change in the change in the 
vacancy formation energy vs Ge concentration from a rule-of-mixtures model
for the composition dependence of the formation energy (dotted line). The
rule-of-mixtures model is consistent with a spatially uniform distribution
of vacancies for each Si$_{1-x}$Ge$_x$ composition.  The difference between the
two curves has a maximum at a particular concentration of Ge, which, of
course, is temperature dependent.  This is understood by the
following reasoning:  At very high Ge concentrations, a randomly chosen
site would have a high probability of having many Ge neighbors.  The
further reduction in the formation energy because of the vacancies
preferentially forming at high Ge concentration sites is therefore
marginal.  At very low Ge concentrations, the amount of reduction in the
formation energy is low because of the small number of Ge atoms present. 

\subsection{\label{sec:vacDB}Vacancy migration energy barrier database}

\begin{figure}
\includegraphics[width=8.5cm]{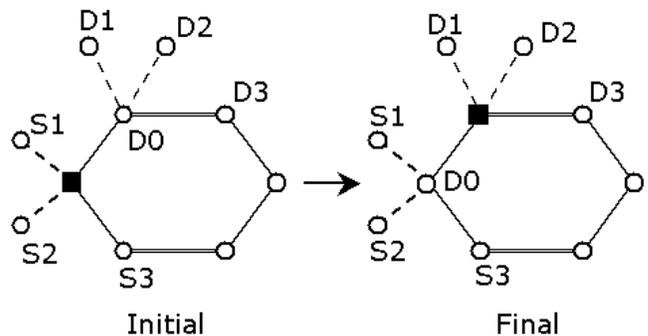}
\caption{\label{fig:vacmig1nn} 
The energy barrier for the vacancy (filled square) to to go from
  the initial configuration to the final is influenced the most by the
  identities of the atoms surrounding the vacancy (S1, S2, S3),
  the identity of the atom with which the vacancy is to exchange position
  (D0), and the identities of the atoms surrounding D0, namely,
  D1, D2, and D3. A two dimensional representation of the diamond
  structure has been adopted for convenience with the different types of
  lines representing bonds on different planes.}
\end{figure}

We present, in this subsection, the database of energy barriers for vacancy
migration in different environments calculated using the local density
approximation. As in the case of calculating the effective vacancy
formation energy, we have treated atoms at different distances from the
vacancy migration center differently depending on the extent of influence
that the atom would exert on the vacancy migration energy barrier.
Referring to Fig.~\ref{fig:vacmig1nn}, the identities of 
the atoms that are first nearest neighbors to the vacancy (denoted as S1,
S2, S3 in the figure), the identity of the migrating atom (denoted as D0),
and the identities of the atoms surrounding the migrating atom (denoted as
D1, D2, D3) are expected to have the greatest influence on the migration
energy barrier. We have assumed that the concentration of vacancies is
sufficiently low that none of the seven atoms (S1 - S3, D0 - D3) would
be a vacancy. We then get a list of 40 different configurations depending
on which of (S1 - S3, D0 - D3) is Si or Ge. We account for the effect of
the identities of the atoms beyond these seven nearest neighbors in the
following mean field manner: We calculate the migration energy barriers
for the 40 different configurations for the following two cases: (i) All
the atoms 
beyond the seven nearest neighbors are Si. (ii) All the atoms beyond
the seven nearest neighbors are Ge. Then, to obtain the energy barrier for any
one of these 40 configurations in a Si$_{1-x}$Ge$_x$ alloy (with a Ge
concentration of $x$), we linearly interpolate the migration energy
barrier of that particular configuration from cases (i) and (ii) mentioned
above. 

\begin{figure}
\includegraphics[width=4.7cm]{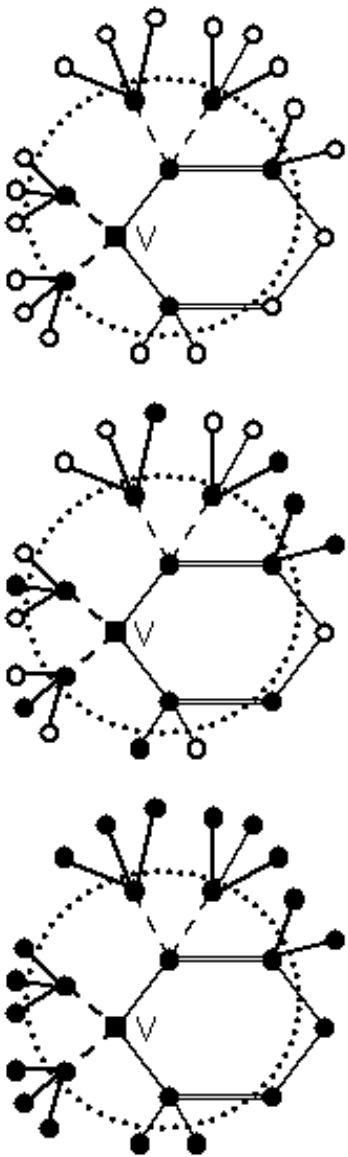}
\caption{\label{fig:vacmig2nn} 
Although all the three configurations shown above have the same
  nearest neighbor atoms to the vacancy migration center (atoms within the
  dotted circle), they have different vacancy migration energy
  barriers. The top configuration has Si atoms (open circles) everywhere
  outside the dotted circle and has a vacancy migration energy barrier of
  0.03 eV. The bottom configuration has Ge atoms (filled circles)
  everywhere outside the dotted circle and has a vacancy migration energy
  barrier of 0.13 eV. The middle configuration has 50\% of the atoms
  outside the dotted circle as Si. The barrier for the forward migration
  is 0.11 eV and that for the reverse migration is 0.07 eV. The mean (0.09
  eV) is closer to the linear interpolation of the barriers from the top
  and the bottom configurations (0.08 eV) than it is to either of
  them. (Si: open circles; Ge: filled circles; vacancy: filled square with
  a label ``V''.)}
\end{figure}

This approach seems to be reasonably satisfactory for at least one
of the configurations (all (S1 - S3, D0 - D3) are Ge atoms) that we have
tested (Fig.~\ref{fig:vacmig2nn}). The top configuration shown in
Fig.~\ref{fig:vacmig2nn} corresponds to the case where the seven nearest
neighbors within the dotted circle (S1 - S3, D0 - D3) are all Ge atoms and
all atoms beyond the 
nearest neighbor sites are Si (case (i) above). The
migration energy barrier is 0.03 eV. The bottom configuration
corresponds to the same nearest neighbor configuration but all atoms
beyond the nearest neighbor sites being Ge (case (ii) above). The
migration energy barrier is 0.13 eV. The middle configuration is an
explicit calculation of the energy barrier with the same nearest neighbor
configuration (i.e., all (S1 - S3, D0 - D3) are Ge atoms) but all the
atoms beyond the nearest neighbor sites are either Si or Ge with a
probability of 0.5. This is consistent with what would occur in a 
Si$_{1-x}$Ge$_x$ alloy with $x$ = 0.5.  Unlike in the top and the bottom
configurations, the barrier for the forward migration (0.11 eV) is
different from that for the reverse migration (0.07 eV) because unlike
in the top and the bottom configurations, the identities of the second
nearest neighbor sites are not identical for the forward and the reverse
migrations. The mean of the forward and the reverse barriers (0.09 eV),
however, is closer to the linear interpolation of the barriers from the
top and the bottom configurations (0.08 eV) than it is to either of the
them. We should mention that the Si$_{0.5}$Ge$_{0.5}$ case is one which is
farthest away from the reference cases (cases (i) and (ii)) and so is a
stringent test, in a certain sense, of the approximation used. So a
reasonable agreement in this case definitely indicates that the
approximation used is reasonable. 

We have calculated the saddle point energies for each of these 80
different configurations very accurately using the nudged elastic band
(NEB) method.\cite{NEB} We also note that we have eliminated strain effects by
suitably adjusting the lattice parameter for the number of Si and Ge
atoms for each of the 80 configurations. Table~\ref{tab:vacDB}  summarizes the vacancy
migration energy barriers. The barriers  
are negligibly small for some of the configurations where there is a
significant asymmetry between the initial and the final environments
of the vacancy (in terms of the number of Ge atoms) especially for the
case of 0\% Ge. By plotting the atomic positions, we have found out that
the reason for this behavior is the following: When there is a significant
assymetry, the initial (or the final) structure ``collapses'' to the
other or to some configration intermediate between the two and there is no
barrier to get from the one to the other. We feel that it happens more in
the case of 0\% Ge because the lower lattice constant of Si compared to Ge
facilitates this collapsing more easily than in the case of 100\% Ge. 

\begin{table}[h]
\caption{\label{tab:vacDB}Vacancy migration
  energy barrier database. S1, S2, 
  and S3 are the identities of the atoms surrounding the vacancy, D0
  is the identity of the atom with which the vacancy is to exchange
  positions, and D1, D2, and D3 are the identities of the atoms
  surrounding D0. Under 0\% Ge are listed the energy barriers
  corresponding to the case where all atoms other 
  than S1 - S3, D0 - D3 are Si and under 100\% Ge 
  are listed the barriers when those atoms are all Ge instead.}
\begin{ruledtabular}
\begin{tabular}{ccccccccc}
 S1 & S2 & S3 & D0 & D1 & D2 & D3 & 0\% Ge
  (eV) & 100\% Ge (eV)\\
\hline
Si & Si & Si & Si & Si & Si & Si & 0.11 & 0.26 \\
Si & Si & Si & Ge & Si & Si & Si & 0.08 & 0.22 \\
Si & Si & Si & Si & Si & Si & Ge & 0.03 & 0.87 \\
Si & Si & Si & Ge & Si & Si & Ge & 0.00 & 0.10 \\
Si & Si & Si & Si & Ge & Ge & Si & 0.00 & 0.05 \\
Si & Si & Si & Ge & Ge & Ge & Si & 0.00 & 0.03 \\
Si & Si & Si & Si & Ge & Ge & Ge & 0.00 & 0.00 \\
Si & Si & Si & Ge & Ge & Ge & Ge & 0.00 & 0.00 \\
Si & Si & Ge & Si & Si & Si & Si & 0.23 & 1.03 \\
Si & Si & Ge & Ge & Si & Si & Si & 0.18 & 0.81 \\
Si & Si & Ge & Si & Si & Si & Ge & 0.09 & 0.86 \\
Si & Si & Ge & Si & Ge & Si & Si & 0.10 & 0.89 \\
Si & Si & Ge & Ge & Si & Si & Ge & 0.06 & 0.18 \\
Si & Si & Ge & Ge & Ge & Si & Si & 0.06 & 0.67 \\
Si & Si & Ge & Si & Ge & Ge & Si & 0.01 & 0.74 \\
Si & Si & Ge & Si & Si & Ge & Ge & 0.00 & 0.70 \\
Si & Si & Ge & Ge & Ge & Ge & Si & 0.00 & 0.09 \\
Si & Si & Ge & Ge & Si & Ge & Ge & 0.00 & 0.07 \\
Si & Si & Ge & Si & Ge & Ge & Ge & 0.00 & 0.64 \\
Si & Si & Ge & Ge & Ge & Ge & Ge & 0.00 & 0.01 \\
Si & Ge & Ge & Si & Si & Si & Si & 0.00 & 0.51 \\
Si & Ge & Ge & Ge & Si & Si & Si & 0.00 & 0.44 \\
Ge & Ge & Si & Si & Si & Si & Ge & 0.21 & 0.34 \\
Ge & Ge & Si & Si & Ge & Si & Si & 0.00 & 0.32 \\
Ge & Ge & Si & Ge & Si & Si & Ge & 0.16 & 0.29 \\
Ge & Ge & Si & Ge & Ge & Si & Si & 0.00 & 0.27 \\
Ge & Ge & Si & Si & Ge & Ge & Si & 0.07 & 0.18 \\
Ge & Ge & Si & Si & Si & Ge & Ge & 0.08 & 0.20 \\
Ge & Ge & Si & Ge & Ge & Ge & Si & 0.04 & 0.15 \\
Ge & Ge & Si & Ge & Si & Ge & Ge & 0.05 & 0.16 \\
Si & Ge & Ge & Si & Ge & Ge & Ge & 0.00 & 0.08 \\
Si & Ge & Ge & Ge & Ge & Ge & Ge & 0.00 & 0.05 \\
Ge & Ge & Ge & Si & Si & Si & Si & 0.00 & 0.69 \\
Ge & Ge & Ge & Ge & Si & Si & Si & 0.00 & 0.00 \\
Ge & Ge & Ge & Si & Si & Si & Ge & 0.00 & 0.48 \\
Ge & Ge & Ge & Ge & Si & Si & Ge & 0.00 & 0.41 \\
Ge & Ge & Ge & Si & Si & Ge & Ge & 0.00 & 0.30 \\
Ge & Ge & Ge & Ge & Si & Ge & Ge & 0.00 & 0.25 \\
Ge & Ge & Ge & Si & Ge & Ge & Ge & 0.05 & 0.16 \\
Ge & Ge & Ge & Ge & Ge & Ge & Ge & 0.03 & 0.13 \\
\end{tabular}
\end{ruledtabular}
\end{table}

\subsection{\label{sec:kmcResults}Kinetic Monte Carlo simulation}

In this section we present the results of the kinetic Monte Carlo
simulations done using the vacancy migration energy barrier database presented
in Sec.~\ref{sec:vacDB}. 

\subsubsection{\label{sec:AMFEresult}Diffusivity, AMF energy}
\begin{figure}
\includegraphics[width=8.5cm]{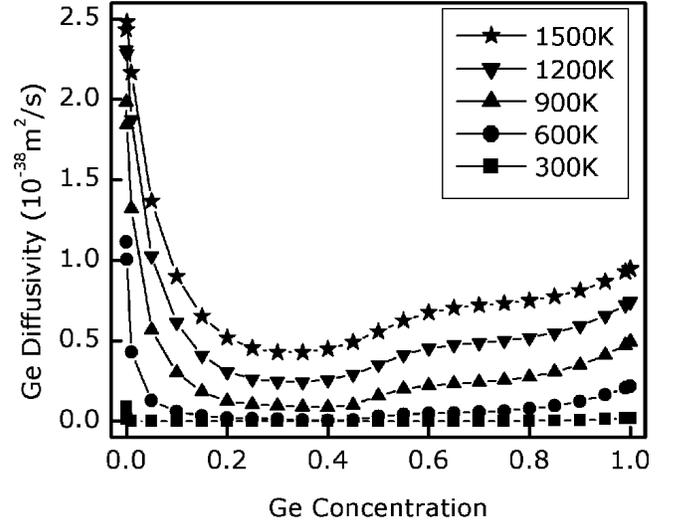}
\caption{\label{fig:GeDiff} 
Diffusivity of Ge in Si$_{1-x}$Ge$_x$ calculated from the results
  of the KMC simulation as a function of Ge concentration ($x$) at
  five different temperatures: 300K - square; 600K - circle; 900K - upward
  triangle; 1200K - downward triangle; 1500K - pentagram. (Lines are
  drawn as a guide to the eye.)}
\end{figure}

\begin{figure}
\includegraphics[width=8.5cm]{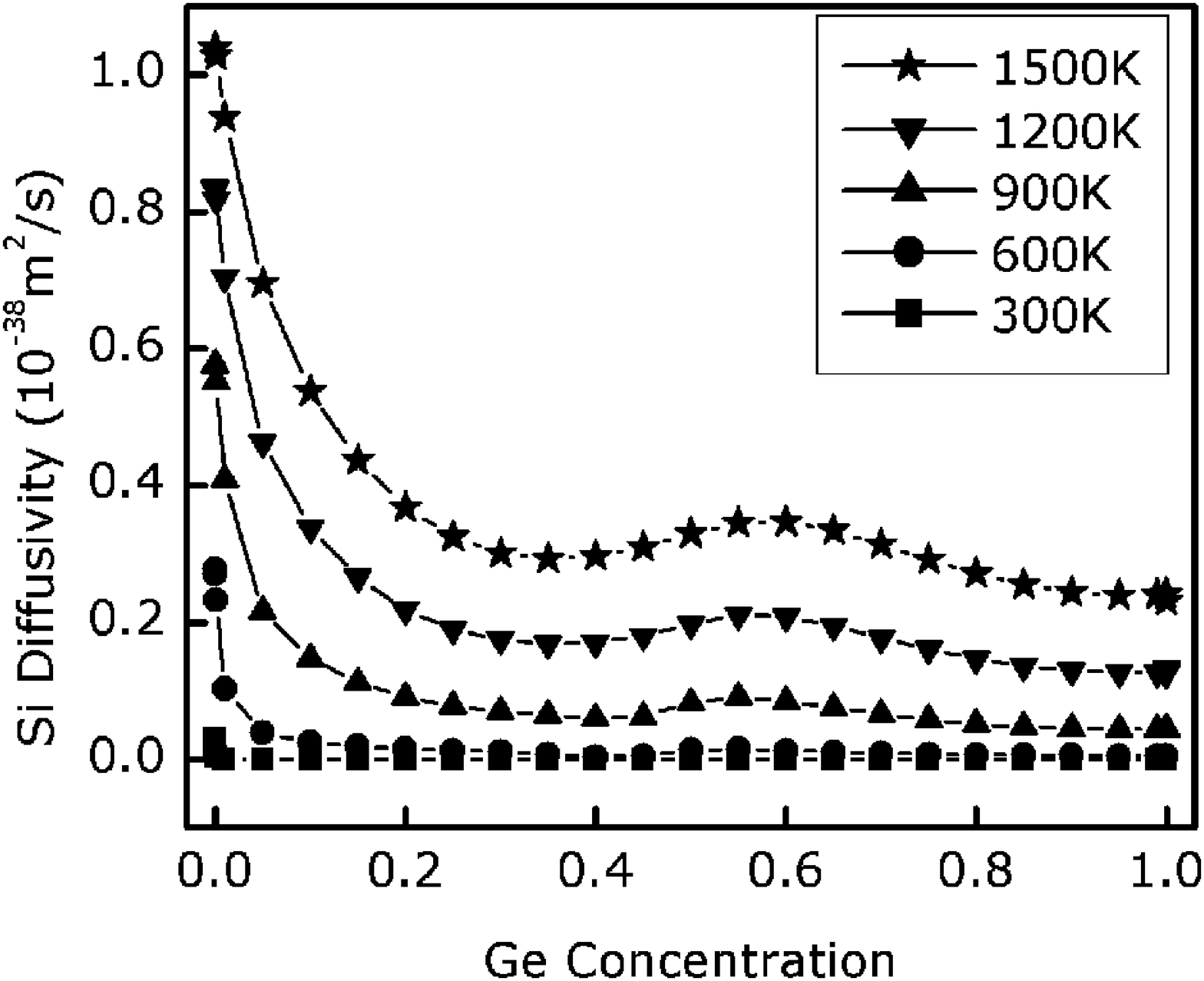}
\caption{\label{fig:SiDiff} 
Diffusivity of Si in Si$_{1-x}$Ge$_x$ calculated from the results
  of the KMC simulation as a function of Ge concentration ($x$) at
  five different temperatures: 300K - square; 600K - circle; 900K - upward
  triangle; 1200K - downward triangle; 1500K - pentagram. (Lines are
  drawn as a guide to the eye.)}
\end{figure}

\begin{figure}
\includegraphics[width=8.5cm]{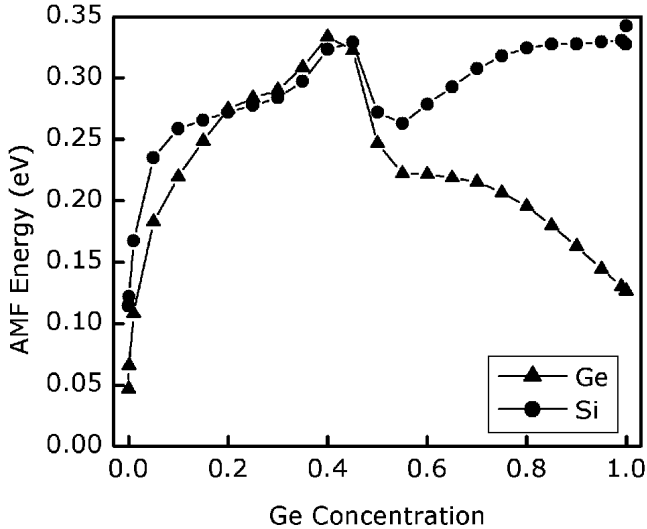}
\caption{\label{fig:GeSiAMFE} 
Variation of the activation-minus-formation (AMF) energy (eV) for
  the diffusion of Ge (upward triangle) and Si (circle) in
  Si$_{1-x}$Ge$_x$ as a function of Ge concentration ($x$) obtained from
  the results of the KMC simulation. (Lines are drawn as a guide to the
  eye.)}
\end{figure}

\begin{figure}
\includegraphics[width=8.5cm]{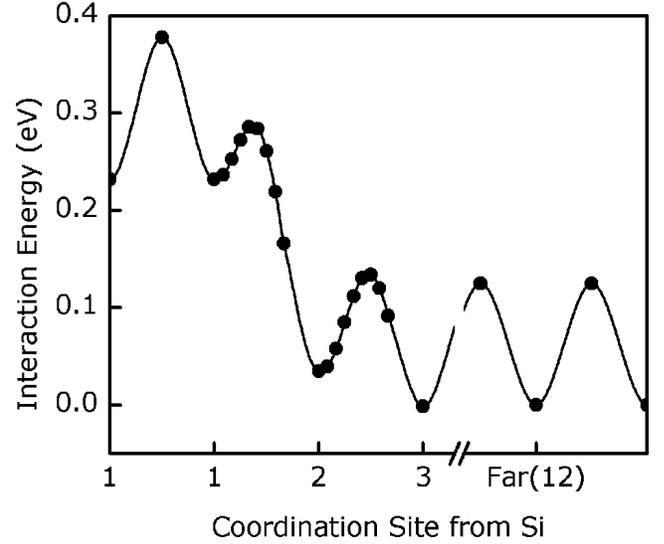}
\caption{\label{fig:SiV} 
Interaction potential energy (in eV) between a substitutional Si
  atom and a vacancy as a function of vacancy position in relaxed Ge from
  LDA calculations. (Lines are drawn as a guide to the eye.)}
\end{figure}

From the KMC simulations, we were able to compute the diffusivity of Si
and Ge in Si$_{1-x}$Ge$_x$ as a function of Ge concentration ($x$) and the
temperature. 
(The diffusivity is given by $D = \langle X^2 \rangle/6\tau$, see
Sec.~\ref{sec:CorrKMCTheo}).  We note that 
these simulations have a constant vacancy concentration (of 10$^{-6}$/atom);
in other words, the change in the vacancy concentration due to the change
in the Ge concentration as explained in Sec.~\ref{sec:dEfXGeRslt} has not been
factored in. The lattice vibrational frequency $\nu_0$ was estimated from
first principles based on a harmonic approximation to be 
$7.325 \times 10^{11}$ sec$^{-1}$.  Figures ~\ref{fig:GeDiff} and
~\ref{fig:SiDiff} show the 
plots of the Ge and the Si diffusivities respectively. 
As expected, the diffusivity increases with temperature.  From an
Arrhenius type plot of the diffusivities, we have extracted the activation
minus formation (AMF) energy.  These have
been plotted in Fig.~\ref{fig:GeSiAMFE}.  We make the following observations
with reference to Figs.~\ref{fig:GeDiff} - ~\ref{fig:GeSiAMFE}. (i) The AMF
energy for the diffusion of Si in Si (0.11 eV) matches closely with the
migration energy for a vacancy in pure Si. (Compare with the entry in Table~\ref{tab:vacDB}
corresponding to all (S1 - S3, D0 - D3) being Si under 0\% Ge.) A similar
close match is also obtained for the AMF energy for the diffusion of Ge in Ge
(0.13 eV).(Compare with the entry in table 1 corresponding to all (S1 -
S3, D0 - D3) being Ge under 100\% Ge.) Along with providing a verification
of our 
computer simulation programs, it also corroborates our concept of the AMF
energy as explained in Sec.~\ref{sec:intro}. (ii) We have not been able to
obtain satisfactory numerical agreement of the AMF energy for the
diffusion of Si in Ge (0.34 eV) or for that of diffusion of Ge in Si 
(0.05 eV) based on the models presented in Refs.~\onlinecite{dunh95a,
pank97a, sugi92a}. We however have the following plausible qualitative
explanation: The {\em attractive} interaction between vacancies and Ge atoms
(see Fig.~\ref{fig:GeV}) causes a vacancy to be {\em more available} near the
vicinity of a Ge atom 
to facilitate its diffusion. This probably results in  {\em lowering} the
AMF energy for the Ge diffusion in Si compared to the migration barrier for a
Ge-vacancy exchange 
process in Si (0.08 eV (see the second entry under 0\% Ge in Table
~\ref{tab:vacDB})). Conversely, the {\em repulsive} interaction between a Si and a vacancy
(see Fig.~\ref{fig:SiV}) makes a vacancy {\em less available} near the
vicinity of a Si atom. This 
probably results in {\em increasing} the AMF energy for the Si diffusion
in Ge compared to the migration barrier for a Si-vacancy exchange process in Ge
(0.16 eV (see the penultimate entry under 100\% Ge in Table ~\ref{tab:vacDB})). (iii)
While we do not have a microscopic explanation for the abrupt drop in the
diffusivity of both Si and Ge near low Ge concentrations, we do find them
to be consistent with the rise in AMF energy of both Si and Ge near low Ge
concentrations. (iv) The reason for the non smooth behavior of the AMF
energies near 50\% Ge concentration is probably because of those
concentrations being farthest away from the reference configurations (0\%
and 100\% Ge) that were used to build the migration energy barrier database.

\begin{figure}
\includegraphics[width=8.5cm]{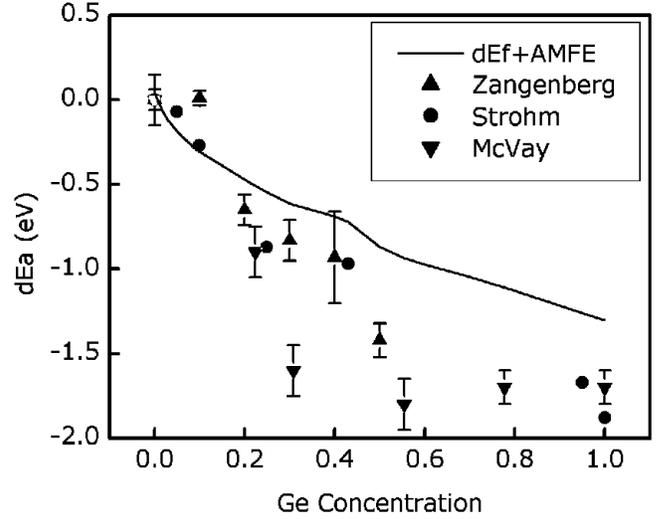}
\caption{\label{fig:dEaALL} 
Solid line shows the change in the activation energy for the
  diffusion of Ge in Si$_{1-x}$Ge$_x$ by a vacancy mechanism from that in
  pure Si as a function of Ge concentration ($x$) calculated at 1000K as
  the sum of the change in the vacancy formation energy and the Ge
  activation-minus-formation (AMF) energy. Also shown are the experimental
  results for the same quantity from Zangenberg {\em et al.}\protect\cite{zang01a}
  (upward triangle), Strohm {\em et al.}\protect\cite{stro01a} (circle), and McVay
  and DuCharme\protect\cite{mcva74a} (downward triangle).} 
\end{figure}

In Fig.~\ref{fig:dEaALL}, we plot the change in the activation energy for
Ge diffusion in Si$_{1-x}$Ge$_x$ compared to that in Si as a function of
Ge concentration. The activation energy is calculated as a sum of the
vacancy formation energy from Fig.~\ref{fig:dEfTheo} and the Ge AMF energy
from Fig.~\ref{fig:GeSiAMFE}. Also plotted on the same axes are the
experimentally observed changes in the activation energy for Ge diffusion
from Refs.~\onlinecite{zang01a,stro01a,mcva74a}. The purpose of this plot
is not actually to compare our results with the experiments, which, as we
have already mentioned in Sec.~\ref{sec:intro} is premature at this
stage. But this plot clearly brings out the need to consider the other
mechanisms before a fair comparison with experiments is possible.

\subsubsection{\label{sec:fcorrResult}Correlation factor}
\begin{figure}
\includegraphics[width=8.5cm]{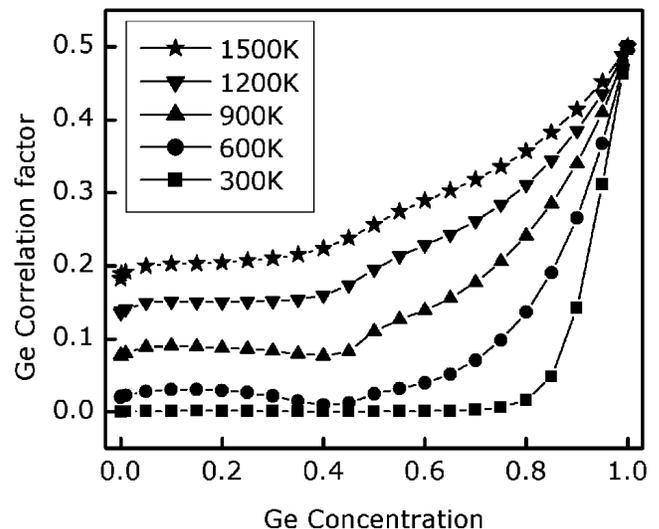}
\caption{\label{fig:GeCorr} 
Correlation factor for the diffusion of Ge in Si$_{1-x}$Ge$_x$
  calculated from the results of the KMC simulation as a function of Ge
  concentration ($x$) at five different temperatures: 300K - square; 600K
  - circle; 900K - upward triangle; 1200K - downward triangle; 1500K -
  pentagram. (Lines are drawn as a guide to the eye.)}
\end{figure}

\begin{figure}
\includegraphics[width=8.5cm]{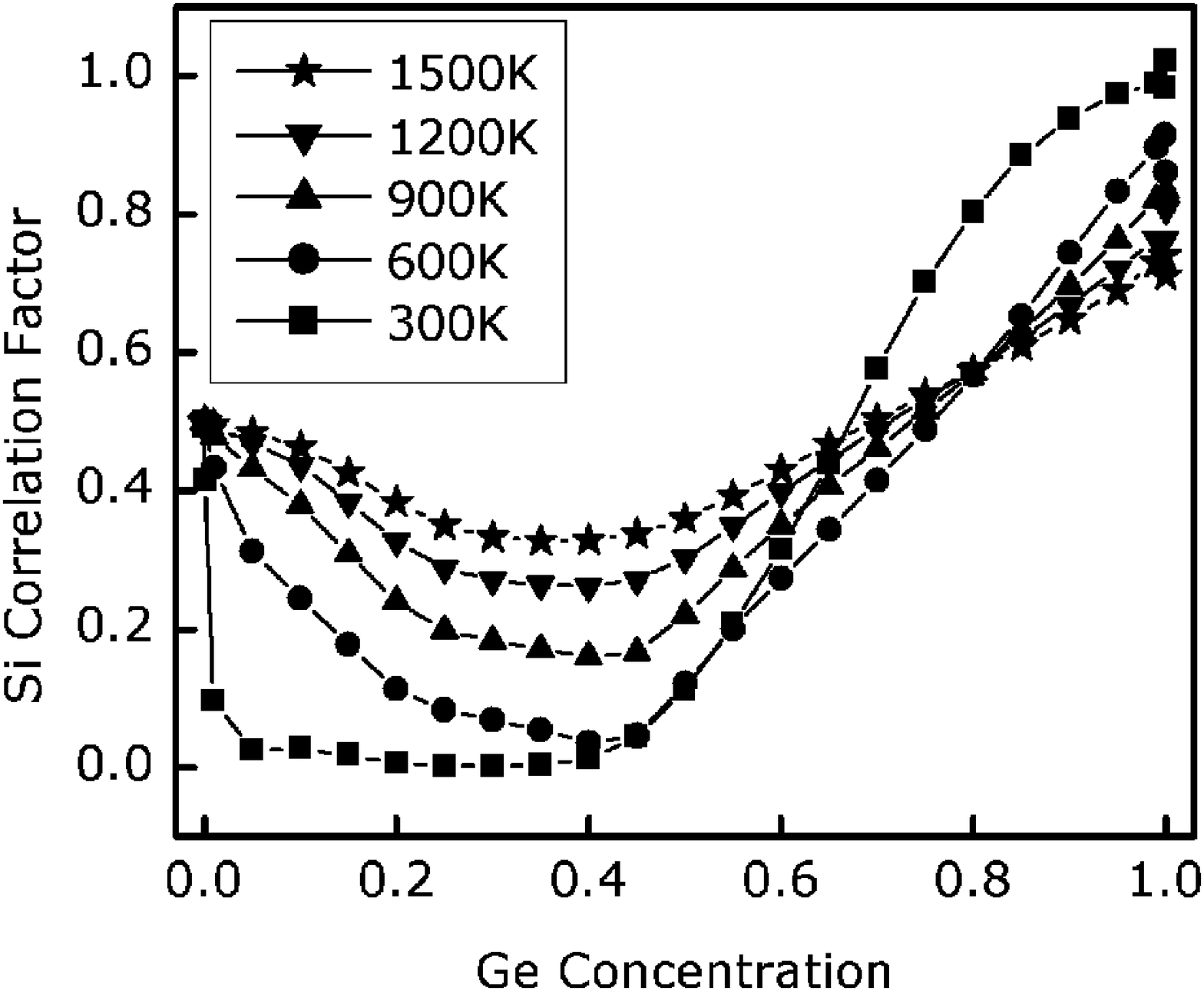}
\caption{\label{fig:SiCorr} 
Correlation factor for the diffusion of Si in Si$_{1-x}$Ge$_x$
  calculated from the results of the KMC simulation as a function of Ge
  concentration ($x$) at five different temperatures: 300K - square; 600K
  - circle; 900K - upward triangle; 1200K - downward triangle; 1500K -
  pentagram. (Lines are drawn as a guide to the eye.)}
\end{figure}

\begin{figure}
\includegraphics[width=8.5cm]{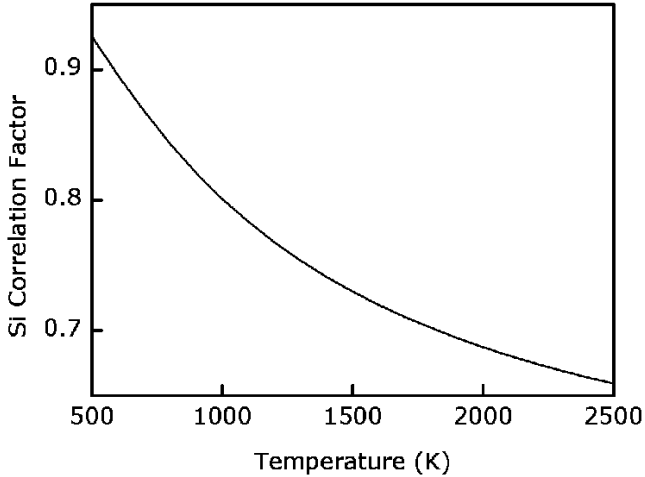}
\caption{\label{fig:SiCorrTheo} 
Theoretical calculation of the correlation factor for the
  diffusion of Si in relaxed Ge.}
\end{figure}

The correlation factor for Ge and Si calculated by the procedures outlined
in Secs.~\ref{sec:CorrTheo} and \ref{sec:CorrKMCTheo} are plotted
respectively in Figs.~\ref{fig:GeCorr} and \ref{fig:SiCorr} as a function
of the Ge concentration for five different temperatures.  
We make the following 
observations with reference to these plots.  (i) The correlation factors
that we have 
calculated for the unary substances (i.e., Si correlation factor in 0\% Ge
concentration and Ge correlation factor in 100\% Ge concentration) equal
0.5.  This is the theoretical value for the correlation factor for a
tracer diffusion in the diamond structure.\cite{mann68a}  (This provides an
additional verification of our calculations.)  That the correlation factor
for a tracer (diffusing by the vacancy mechanism) should be less than
unity is clear from the observation that the tracer atom has a higher
probability of jumping back to the vacancy site thereby nullifying a
forward jump.  The mean squared displacement (and hence the diffusivity)
of this correlated motion would therefore be less than that of a random
jump, giving a correlation factor less than unity.  The specific value of
0.5 is a result of the tetrahedral geometry of the silicon crystal structure.
(ii) At higher temperatures, the Boltzmann factor evens out the different
energy barriers, making the system resemble a unary substance.  One would
therefore expect the correlation factor to approach the value for a unary
substance in the diamond structure, namely 0.5.  We observe this in our
plots.  (iii) At low Ge concentrations, the
correlation factor for Ge drops below 0.5.  The reason for
this is understood by the following argument:  The attractive
interaction between Ge and a vacancy and the lower energy barrier for a
vacancy to exchange positions with Ge than to jump to the second nearest
neighbor site of Ge from the first (which results in breaking the
Ge-vacancy bond, see Fig.~\ref{fig:GeV}), tend
to cause the Ge and the vacancy to jump back and 
forth several times before breaking away from each other. But, in the diamond
structure, because there is no atomic location that is a simultaneous
neighbor to both the vacancy and the Ge atom (when they are first nearest
neighbors to each other), breakage of the Ge-vacancy pair is essential for
the Ge atom to be effectively displaced from its current location.  This
back and forth motion does not contribute to the mean squared displacement
of the Ge atoms and 
consequently the Ge correlation factor drops. (iv) At low Ge
concentrations, the 
correlation factor for Si drops below 0.5. We offer the following
explanation for this behavior: The attractive interaction
between the Ge and the vacancy causes the vacancies to be predominantly
found near Ge atoms. So, only the Si atoms found near those Ge atoms are
affected by the vacancy motion. These Si atoms, owing to the lower energy
barrier for the vacancy to jump to the first nearest neighbor site of the
Ge atom from the second than to jump to the third from the second (see
Fig.~\ref{fig:GeV}), just
keep jumping back and forth between the first and the second nearest
neighbor sites of the Ge (depending on whether the vacancy is
correspondingly at the second or the first nearest neighbor sites). This
back and forth motion does not effectively displace the Si atoms and so
does not contribute to the mean squared displacement. This causes the Si
correlation factor to drop.
(v) The correlation factors of both
Si and Ge increase with increasing Ge concentration.  We explain this in
the following manner:  As the concentration of Ge increases, the vacancy
is attracted by the other Ge atoms too and therefore it is less likely to
be bound to a single Ge atom. This reduces the
redundant back and forth motion of 
the Ge atoms, thus increasing the mean squared displacement and
consequently the Ge correlation factor.  (This effect is similar, in some
ways, to the percolation mechanism for diffusion.\cite{math82a}) 
The vacancy, in the process of moving from one Ge atom to the other, ends
up displacing Si atoms thus increasing their mean squared displacement and
consequently the Si correlation factor. (vi) The correlation factor for Si
in Si$_{1-x}$Ge$_x$ alloys with high 
Ge concentration is greater than 0.5 and approaches unity.  This
interesting behavior is explained by the following reasoning:  At very
high Ge concentration (i.e., very low Si concentration), the faster
jumping rate of the vacancy with the Ge atoms compared to that with the Si
atoms (because of the 
lower barrier height (compare last two entries under 100\% Ge in Table ~\ref{tab:vacDB}))
causes the vacancy to perform a lot of jumps with Ge atoms between
successive jumps with a Si atom. This results in the vacancy approaching
Si via an essentially random path, making the Si jumps closer to a random walk
process. This causes the correlation factor to approach unity.
In Fig.~\ref{fig:SiV}, we show the interaction energy between
a Si and 
a vacancy in a Ge environment.  In Fig.~\ref{fig:SiCorrTheo} we show the
variation of the Si
correlation factor with temperature calculated as outlined in
Sec.~\ref{sec:CorrTheo}.  We do see that the correlation factor tends to
unity in the lower temperature limit. 

\section{\label{sec:summary}Summary}
Our purpose of the present work was to understand, from first principles,
the effect of biaxial strain and composition on the self-diffusivity of Si
and Ge in Si$_{1-x}$Ge$_x$ alloys.  In order to attack the problem, we broke it
down into one of studying the effect of these factors on the main
components that define the diffusivity: the vacancy formation energy, and
the activation minus formation (AMF) energy.  (The necessity and the
definition of AMF energy were
presented.)  We attacked the problem by the following three
main steps:  (i) We performed density functional theory (DFT) calculations
within the local density approximation (LDA) to obtain the required
energetics of the various configurations.  (ii) We
worked out the details necessary to calculate the correlation factor and
the change in the vacancy formation energy with composition.  (iii)
We performed kinetic Monte Carlo (KMC) simulations using our
total energy calculations.  By this approach, we were able to estimate the
following values for the effect of biaxial strain on the activation energy
(the sum of the vacancy formation energy and AMF energy): 20 eV/unit
strain for Si self diffusion in Si and 17 - 20 eV/unit strain for Ge
self-diffusion in Si.
We calculated the change in the vacancy formation energy in 
Si$_{1-x}$Ge$_x$ as a function of composition.  From the KMC simulations,
we were 
able to extract the variation of the AMF energy 
for Si and Ge self-diffusion in Si$_{1-x}$Ge$_x$ as a function of
composition.  We 
combined the Ge AMF energy with the vacancy formation energy to find the
variation of the activation energy for Ge diffusion in Si$_{1-x}$Ge$_x$ as a
function of composition.  Lastly, we presented the variation of the
correlation factor for Si and Ge diffusion in Si$_{1-x}$Ge$_x$ as a function of
composition and temperature and made several interesting observations that
are quite general for a vacancy mediated diffusion in a random binary
alloy arranged in a diamond structure. 

There are many outstanding issues of the complete model that need to be
resolved even for the vacancy mechanism alone.  We conclude this article
by recognizing the following limitations of the present work:  (i) As we
mentioned in the introduction, the inability of the LDA to reproduce
experimentally observed values of the activation energy in Si precludes
our results from being directly compared with experiments.  
(ii) We have not addressed the effect of strain and composition on the
pre-exponential factor and have not considered entropic effects.

\begin{acknowledgments}
P.R. gratefully acknowledges the feedback from Prof. P. C. McIntyre and
Dr. P. B. Griffin and thanks them for carefully reviewing the manuscript.
P.R. also thanks Dr. R. Sabiryanov, S. Park, G. Lakatos, and D. Aubertine
for useful discussions.  The computations were performed on the Cray T3E,
IBM p690 and on {\em Multipod}, a PC cluster of MSL, Stanford.
Utilization of the Cray and the IBM machines were possible through a grant
from NPACI. Efforts of G. Jun in maintaining {\em Multipod} are sincerely
appreciated. This work was funded by the Department of Energy, Basic Energy
Sciences grant DE-FG03-99ER45788.
\end{acknowledgments}

\bibliography{SiGelanl}

\begin{thebibliography}{55}
\bibitem[{mur()}]{murnFN}
\bibinfo{note}{The Murnaghan's equation of state \cite{murn44a} is a relation
  between the energy $E$ and the volume $V$ of the supercell and has the
  following form: $E = E_0 + [(V-V_0)/b] - \{[(V^{1-a} -
  V_0^{1-a})V_0^a]/[(1-a)b]\}$. $E_0$, $V_0$, $a$, and $b$ are fitting
  parameters.}
\bibitem[{geg()}]{gegeFN}
\bibinfo{note}{Our calculations show that the energies of the systems with 62
  Si atoms and 2 Ge atoms with the Ge atoms being at different positions
  relative to each other differ by less than 0.009 eV confirming that
  Si$_{1-x}$Ge$_x$ does indeed form a model random alloy.}
\bibitem[{cor()}]{correlFN}
\bibinfo{note}{We justify our choice of limiting the number of vacancy hops to
  five based on the following observation: The calculated correlation factors
  at a temperature of $2 \times 10^5$ K considering upto one, three, and five
  vacancy jump sequences respectively are 0.579, 0.518, and 0.509. At such a
  high temperature as $2 \times 10^5$ K, one would expect to get the
  theoretical correlation factor for a tracer diffusion in a diamond lattice
  which has a value \cite{mann68a} of 0.5. Using five hops, we are able to get
  reasonably close to the theoretical value.}
\end{thebibliography}

\end{document}